\documentclass{article}
\usepackage{geometry}
\usepackage{wrapfig}
\usepackage{graphicx}
\usepackage{bm}
\makeatletter
\newcommand{\figcaption}[1]{\def\@captype{figure}\caption{#1}}
\newcommand{\tblcaption}[1]{\def\@captype{table}\caption{#1}}
\makeatother
\newcommand{\lw}[1]{\smash{\lower 1.5ex\hbox{#1}}}
\newcommand{\ri}[1]{\smash{\raise 1.5ex\hbox{#1}}}

\newcommand{\mapright}[1]{\smash{\mathop{\hbox to 1cm{\rightarrowfill}}\limits^{#1}}}

\begin{document}

\title{Unified Analyses of Multiplicity Distributions and Bose--Einstein Correlations at the LHC using Double-Stochastic Distributions}

\author{Takuya \textsc{Mizoguchi}$^{1}$ and Minoru \textsc{Biyajima}$^{2}$\\
{\small $^{1}$National Institute of Technology, Toba College, Toba 517-8501, Japan}\\
{\small $^{2}$Department of Physics, Shinshu University, Matsumoto 390-8621, Japan}}

\date{}
\maketitle

\begin{abstract}
We analyze data on multiplicity distributions (MD) at Large Hadron Collider (LHC) energies using a double-negative binomial distribution (D-NBD) and double-generalized Glauber--Lachs formula (D-GGL). Moreover, we investigate the Bose--Einstein correlation (BEC) formulas based on these distributions and analyze the BEC data using the parameters obtained by analysis of MDs. From these analyses it can be inferred that the D-GGL formula performs as effectively as the D-NBD. Moreover, our results show that the parameters estimated in MD are related to those contained in the BEC formula.
\end{abstract}

\section{\label{sec1}Introduction}
In Large Hadron Collider (LHC) experiments, various multiplicity distributions (MD) have been measured with a pseudorapidity interval ($|\eta| < \eta_c$) in proton-proton collision \cite{Aad:2010ac,Aaboud:2016itf,Khachatryan:2010nk}. Additionally, the negative binomial distribution (NBD) has been used to analyze MD.
\begin{eqnarray}
 P_{\rm NBD}(n,\,k,\,\langle n\rangle) = \frac{\Gamma (n+k)}{\Gamma (n+1)\Gamma (k)}\frac{(\langle n\rangle/k)^n}{(1+\langle n\rangle/k)^{n+k}}.
\label{eq1}
\end{eqnarray}
However, we cannot understand the behavior of the parameter: $k= 7,\, 8$ for lower energy (ISR) and $k = 2,\,3$  for higher energy (S$p\bar p$S).

The double-negative binomial distribution (D-NBD) \cite{Giovannini:1998zb,Ghosh:2012xh} has been proposed to explain the KNO scaling \cite{Koba:1972ng} violation observed by the UA5 \cite{Fuglesang:1989st,Alner:1985zc,Ansorge:1988kn}, using $\alpha_1+\alpha_2=1$,
\begin{eqnarray}
  P_{\rm D\mathchar`-NBD}(n,\, \langle n\rangle) = \alpha_1 P_{\rm NBD_1}(n,\,k_1,\,\langle n_1\rangle) + \alpha_2P_{\rm NBD_2}(n,\,k_2,\,\langle n_2\rangle).
\label{eq2}
\end{eqnarray}

In 2010, we analyzed the ALICE data \cite{Aamodt:2010pp} by the generalized Glauber--Lachs (GGL) formula, and showed that GGL fits as well as NBD \cite{Mizoguchi:2010vc}:
\begin{eqnarray}
P_{\rm GGL}(n,\,k,\,p,\,\langle n\rangle) = \frac{(p\langle n\rangle/k)^n}{(1+p\langle n\rangle/k)^{n+k}}
\exp\left[-\frac{\gamma p\langle n\rangle}{1+p\langle n\rangle/k}\right]
L_n^{(k-1)}\left(-\frac{\gamma k}{1+p\langle n\rangle/k}\right),
\label{eq3}
\end{eqnarray}
where $p = 1/(1 + \gamma)$ and $\gamma$ reflects the contamination ($K$, $p/\pi$'s) and the degree of superposition of the phase spaces of the particles. Our analyses suggested that the coherent component is necessary. Moreover, Eq.~(\ref{eq3}) has the following stochastic property:
\begin{eqnarray}
\mbox{Eq.~(\ref{eq3})}\ \left\{
\begin{array}{l}
\mapright{\quad\ k=1\qquad}\ \mbox{original GL}\ \mapright{\quad\ p\to 1\qquad}\ \mbox{Furry distribution}\medskip\\
\mapright{\quad\ p\to 1\qquad}\ \mbox{NBD of Eq.~(\ref{eq1})}\\
\hspace{30mm} \downarrow k\to \infty\\
\mapright{\quad\ p\to 0\qquad}\ \mbox{Poisson distribution}.
\end{array}
\right.
\label{eq4}
\end{eqnarray}

To explain the shape of the MD with higher energy than S$p\bar p$S, two sources or two collision mechanisms \cite{GrosseOetringhaus:2009kz,Navin:2010kk,Biyajima:2003nv} are required; thus, the two-component model is necessary.

In theoretical analysis of experimental data, MD and the Bose--Einstein correlation (BEC) have been handled as independent observables. In contrast, we have proposed a method for analyzing MD and BEC data using common parameters \cite{Biyajima:1979ak,Biyajima:1990ku, Biyajima:1978cz}. See a review book \cite{Kittel:2005fu}.\\

\section{\label{sec2}Double-GGL formula (D-GGL)}
\begin{figure}[htbp]
\vspace{-0mm}
  \centering
  \includegraphics[width=0.34\columnwidth]{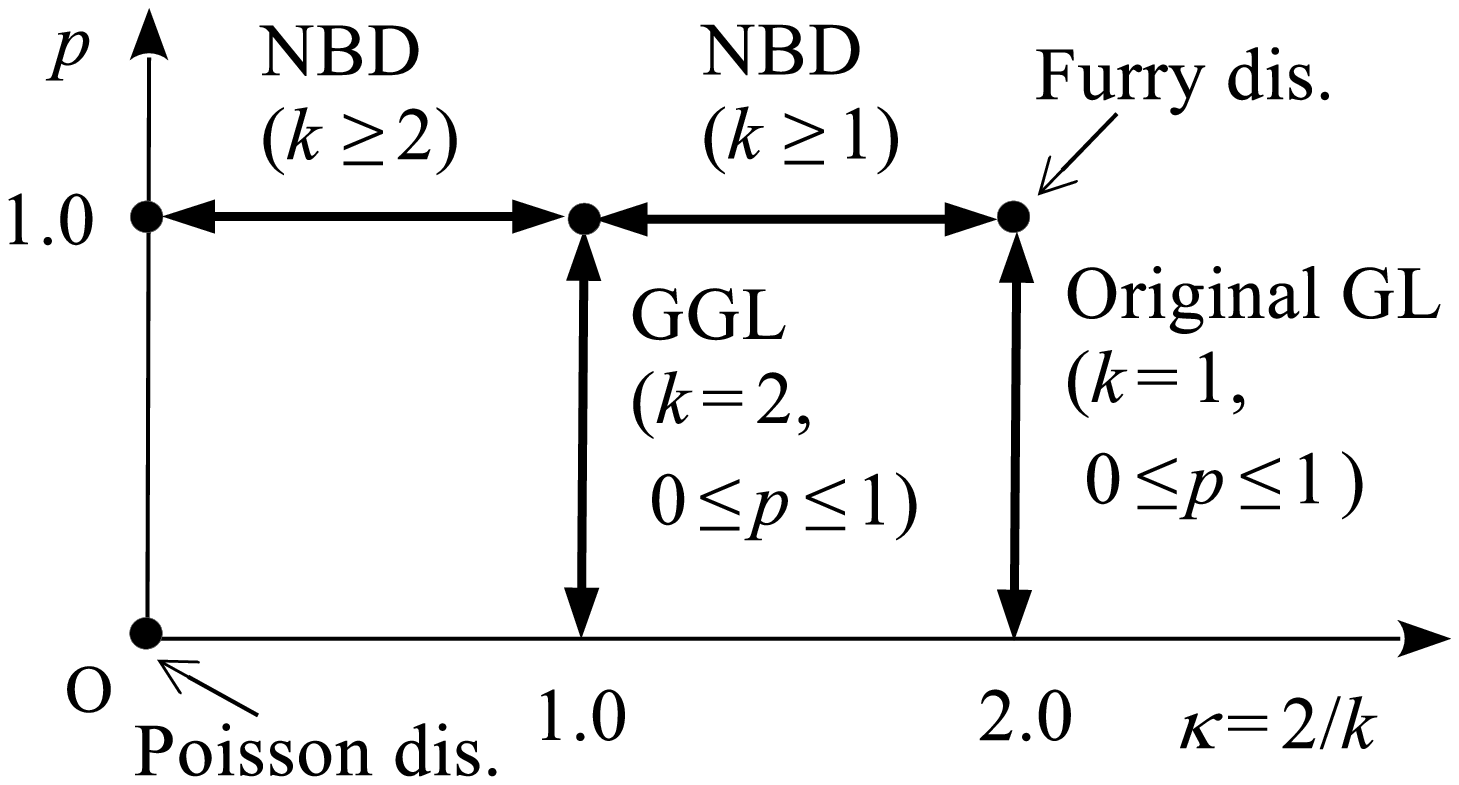}\qquad\qquad
  \includegraphics[width=0.40\columnwidth]{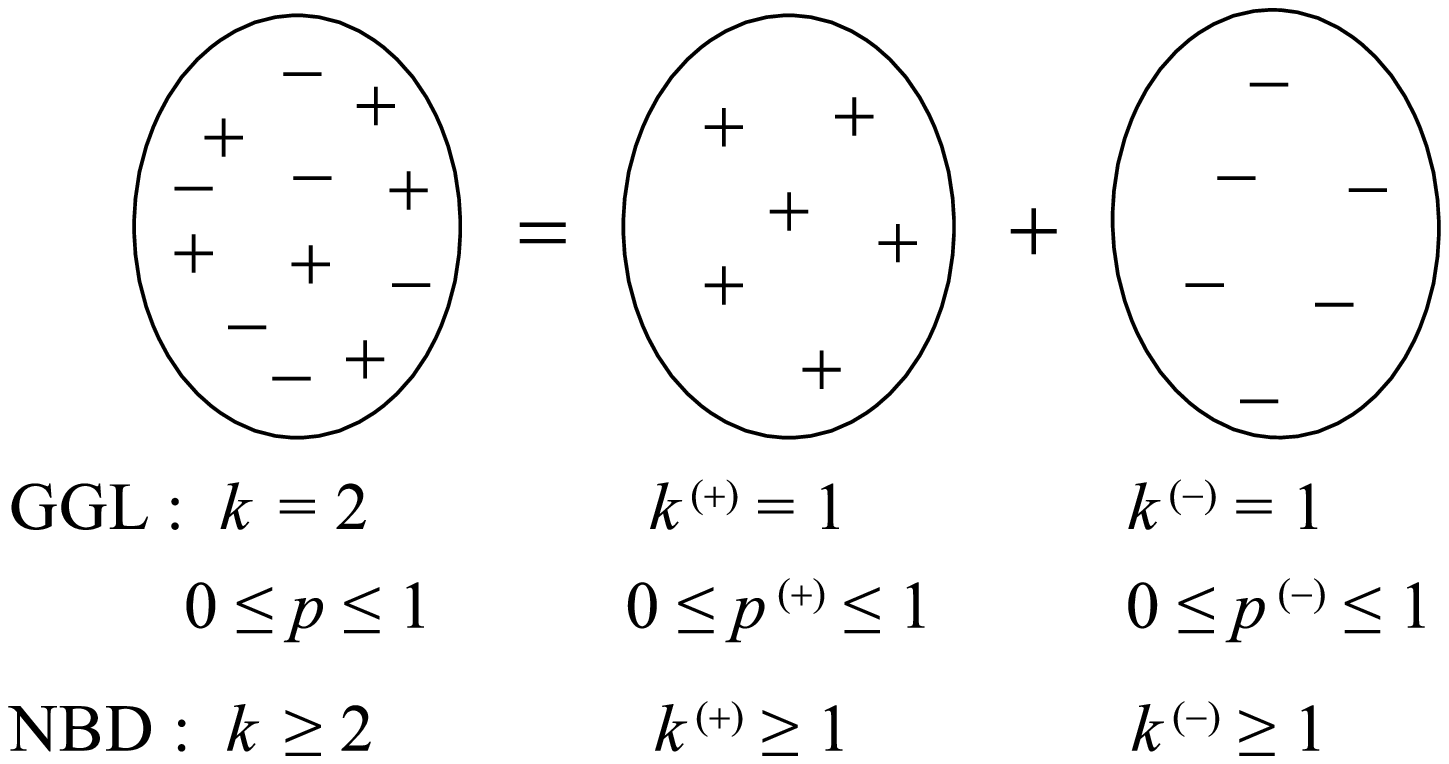}\\
  (a)\hspace{70mm} (b)
  \caption{\label{fig1}(a) The stochastic properties of the generalized Glauber--Lachs formula are shown in the $\kappa$--$p$ plane. At point (2.0, 1.0), the perfect Bose--Einstein statistics for the same charged particles ensemble hold. At point (1.0, 1.0), the Bose--Einstein statistics for the $(+-)$ charged particle ensemble hold. (b) The charged particle ensemble is decomposed into the positive $(+)$ and negative $(-)$ ensembles. Here $\gamma= \langle n_{\rm coherent}\rangle/\langle n_{\rm chaotic}\rangle$ can be assumed to reflect the contamination ($K$, $p$/$\pi$'s) and the degree of superposition of the phase spaces of the particles. }
\vspace{-5mm}
\end{figure}

\begin{figure}[htbp]
  \centering
  \includegraphics[width=0.43\columnwidth]{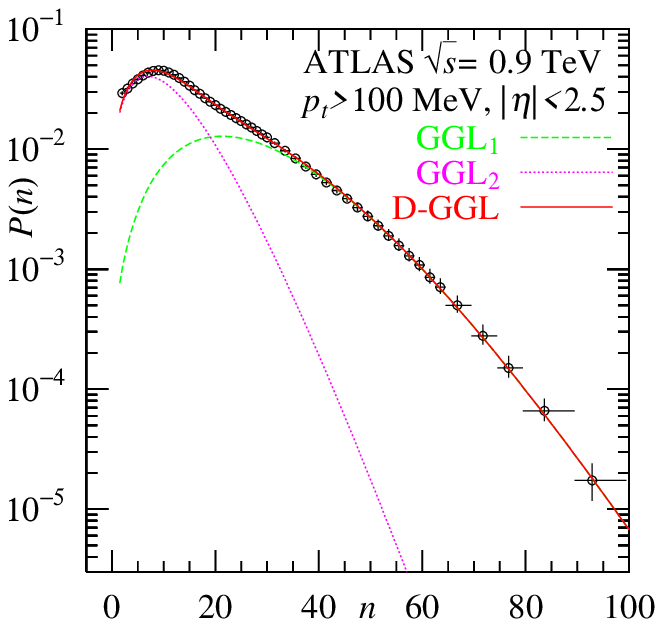}
  \includegraphics[width=0.43\columnwidth]{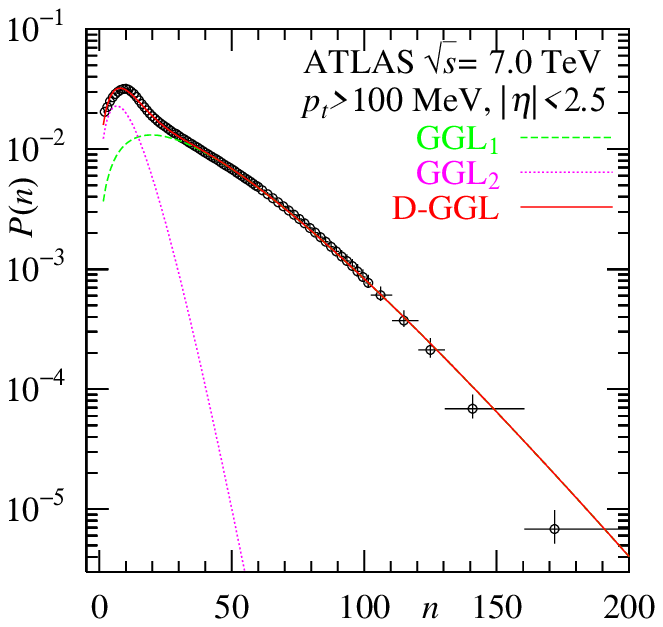}
  \caption{\label{fig2}Analyses of ATLAS data on the MD ($p_t>100$ MeV, $|\eta| < 2.5$, $n_{\rm ch} \ge 2$) using Eq.~(\ref{eq5}).}
\vspace{-5mm}
\end{figure}

Herein, we propose the D-GGL formula\cite{Biyajima:2018abe} for analyzing the multiplicity distribution at LHC:
\begin{eqnarray}
  P_{\rm D\mathchar`-GGL}(n,\, \langle n\rangle) = \alpha_1 P_{\rm GGL_1}(n,\,k=2,\,p_1,\,\langle n_1\rangle) + \alpha_2P_{\rm GGL_2}(n,\,k=2,\,p_2,\,\langle n_2\rangle),
\label{eq5}
\end{eqnarray}
where $k = 2$ reflects the degree of freedom for the $(+-)$ particle ensembles (Fig.~\ref{fig1}). Our results are presented in Fig.~\ref{fig2} and Table \ref{tab1}.

To understand the KNO scaling violation, the weight factors ($\alpha_1$ and $\alpha_2$) are displayed in Fig.~\ref{fig3}. The fluctuation of ($\alpha_1$ and $\alpha_2$) are reflecting the KNO scaling violation.

\begin{table}[htbp]
\centering
\caption{\label{tab1}Analyses results of analyses of ATLAS data on the MD ($p_t>100$ MeV, $|\eta| < 2.5$, $n_{\rm ch} \ge 2$) using Eqs.~(\ref{eq2}) and (\ref{eq5}).} 
\vspace{1mm}
\begin{tabular}{c|cccccc}
\hline
$\sqrt s$ & \multicolumn{6}{c}{Eq.~(\ref{eq2})\quad D-NBD}\\
$[$TeV] & $\alpha_1$ & $k_1$ & $\langle n_1\rangle$ & $k_2$ & $\langle n_2\rangle$ & $\!\!\chi^2/$ndf$\!\!$\\
\hline
0.9$\!\!$
& 0.23$\pm$0.03$\!\!$
& 7.52$\pm$0.77$\!\!$
& 32.9$\pm$1.3$\!\!$
& 2.70$\pm$0.12$\!\!$
& 12.6$\pm$0.4$\!\!$
& 70.2/46$\!\!$\\
7.0$\!\!$
& 0.60$\pm$0.02$\!\!$
& 2.61$\pm$0.10$\!\!$
& 39.4$\pm$0.6$\!\!$
& 2.72$\pm$0.09$\!\!$
& 11.0$\pm$0.1$\!\!$
& 129/80$\!\!$\\
\hline
$\sqrt s$ & \multicolumn{6}{c}{Eq.~(\ref{eq5})\quad D-GGL}\\
$[$TeV] & $\alpha_1$ & $p_1$ & $\langle n_1\rangle$ & $p_2$ & $\langle n_2\rangle$ & $\!\!\chi^2/$ndf$\!\!$\\
\hline
0.9$\!\!$
& 0.41$\pm$0.05$\!\!$
& 0.25$\pm$0.04$\!\!$
& 27.0$\pm$1.4$\!\!$
& 0.39$\pm$0.02$\!\!$
& 10.6$\pm$0.4$\!\!$
& 60.3/46$\!\!$\\
7.0$\!\!$
& 0.67$\pm$0.01$\!\!$
& 0.68$\pm$0.03$\!\!$
& 36.7$\pm$0.3$\!\!$
& 0.42$\pm$0.02$\!\!$
& 10.4$\pm$0.1$\!\!$
& 114/80$\!\!$\\
\hline
\end{tabular}
\end{table}

\section{\label{sec3}Bose--Einstein correlations (BEC)}
By using Eq.~(\ref{eq5}), we obtain the moments as
\begin{eqnarray}
  \langle n\rangle &=& \alpha_1 \langle n_1\rangle + \alpha_2 \langle n_2\rangle,\nonumber\\
  \langle n(n-1)\rangle &=& \alpha_1 \langle n_1(n_1-1)\rangle + \alpha_2 \langle n_2(n_2-1)\rangle.
\label{eq6}
\end{eqnarray}
For BEC, we need the ratio $N^{(2+:\,2-)}/N^{\rm BG}$, where $N$ stands for the number of pairs. The numerator $N^{(2+:\,2-)}$ is the number of the pairs of the same charged particles. $N^{\rm BG}$ is that of different charged particles. There are two possibilities for calculating $N^{\rm BG}$. Whole events can be categorized as two ensembles labeled $\alpha_1$ and $\alpha_2$. Each $N_1^{\rm BG}$ and $N_2^{\rm BG}$ are calculated inside those ensembles: $N_1^{\rm BG} = {}_{\langle n_1^{(+)}\rangle}C_1\times {}_{\langle n_1^{(-)}\rangle}C_1 = \langle n_1^{(+)}\rangle\langle n_1^{(-)}\rangle = \langle n_1\rangle^2$ and $N_2^{\rm BG} = \langle n_2\rangle^2$. This procedure needs the complete separation of two ensembles, for this aim.

Conversely, the second $N_{\rm II}^{\rm BG}$ is computed from two ensembles, such as $N_{\rm II}^{\rm BG}= (\alpha_1\langle n_1\rangle^2+\alpha_2\langle n_2\rangle^2$). The denominator is the sum of numerators of pairs come from the first and second ensembles. Thus, we have two formulas for BEC, which depend on $N_{\rm I}^{\rm BG}$ and $N_{\rm II}^{\rm BG}$.
\begin{eqnarray}
 N^{(2+:\,2-)}/N_{\rm I}^{\rm BG} 
&=& \alpha_1\left[1 + \frac 2{k_1}f(p_1,\,E_{\rm BE_1})\right]
+ \alpha_2\left[1 + \frac 2{k_2}f(p_2,\,E_{\rm BE_2})\right],
\label{eq7}\\
 N^{(2+:\,2-)}/N_{\rm II}^{\rm BG} 
&=& \frac{1}{1+\Gamma}\left[1 + \frac 2{k_1}f(p_1,\,E_{\rm BE_1})\right]
+ \frac{1}{1+\Gamma^{-1}}\left[1 + \frac 2{k_2}f(p_2,\,E_{\rm BE_2})\right],
\label{eq8}
\end{eqnarray}
where, $\Gamma = (\alpha_2\langle n_2\rangle^2)/(\alpha_1\langle n_1\rangle^2)$ and the normalization $(1+\Gamma)^{-1}+(1+\Gamma^{-1})^{-1}=1$, and 
\begin{eqnarray}
f(p,\,E_{\rm BE}) = 2p(1-p)E_{\rm BE}+p^2E_{\rm BE}^2.
\label{eq9}
\end{eqnarray}
The exchange function between the same charged particles is expressed by the exponential function (E, $n=1$) or the Gaussian function (G, $n=2$).
\begin{eqnarray}
E_{\rm BE}^2 = \exp(-(RQ)^n),
\label{eq10}
\end{eqnarray}
where $R$ is the magnitude of the interaction range and $Q=\sqrt{-(p_1-p_2)^2}$ is the momentum transfer. For BEC based on D-NBD, we obtain the formula by adapting $p=1.0$ in Eq.~(\ref{eq9}).

Our results are based on the parameters presented in Table \ref{tab1} and Eqs.~(\ref{eq7})$\sim$(\ref{eq10}). The results are displayed in Table \ref{tab2} and Fig.~\ref{fig4}.\\

\begin{table}[htbp]
\vspace{-5mm}
\centering
\caption{\label{tab2}Analyses results of ATLAS data on BEC ($p_t>100$ MeV, $|\eta| < 2.5$, $n_{\rm ch} \ge 2$) \cite{Aad:2015sja} using Eqs.~(\ref{eq7}), (\ref{eq8}), and (\ref{eq11}). The BEC formulas contain the normalization factor and the long range correlation as $(1+\varepsilon Q)$. The conventional formula (CF$_{\rm I}$) is given as CF$_{\rm I}=1.0+\lambda E_{\rm BE}^2$.}
\vspace{1mm}
\begin{tabular}{c|ccc|ccc|ccc}
\hline
$\!\!\sqrt s\!\!$ & \multicolumn{3}{c|}{Eq.~(\ref{eq7})\quad D-NBD}
& \multicolumn{3}{c|}{Eq.~(\ref{eq7})\quad D-GGL}
& \multicolumn{3}{c}{CF$_{\rm I}$}\\

$\!\!\!\!$\scriptsize{[TeV]}$\!\!\!\!$ & $\!\!R_1$ [fm] & $\!\!R_2$ [fm] & $\!\!\!\chi^2/$ndf$\!\!$
&  $\!\!R_1$ [fm] & $\!\!R_2$ [fm] & $\!\!\!\chi^2\!\!$
& $R$ [fm] & $\lambda$ & $\!\!\!\chi^2\!\!$\\

\hline
  0.9
& $\!\!$1.7$\pm$0.4 (E)
& $\!\!$1.7$\pm$0.1 (E)
& $\!\!\!$98.6/75$\!\!$
& $\!\!$2.8$\pm$0.2 (E)
& $\!\!$2.7$\pm$0.1 (E)
& $\!\!\!$148$\!\!$
& $\!\!$1.8$\pm$0.1(E)$\!\!$
& $\!\!$0.62$\pm$0.01$\!\!$
& $\!\!\!$86.0$\!\!$ \\
  7.0
& $\!\!$1.8$\pm$0.0 (E)
& $\!\!$3.1$\pm$0.1 (E)
& $\!\!\!$743/75$\!\!$
& $\!\!$3.6$\pm$0.1 (E)
& $\!\!$3.2$\pm$0.1 (E)
& $\!\!\!$629$\!\!$
& $\!\!$2.1$\pm$0.0(E)$\!\!$
& $\!\!$0.62$\pm$0.01$\!\!$
& $\!\!\!$919$\!\!$ \\
\hline
$\!\!\sqrt s\!\!$ & \multicolumn{3}{c|}{Eq.~(\ref{eq8})\quad D-NBD}
& \multicolumn{3}{c|}{Eq.~(\ref{eq8})\quad D-GGL}\\

$\!\!\!\!$\scriptsize{[TeV]}$\!\!\!\!$ & $\!\!R_1$ [fm] & $\!\!R_2$ [fm] & $\!\!\!\chi^2/$ndf$\!\!$
& $\!\!R_1$ [fm] & $\!\!R_2$ [fm] & $\!\!\!\chi^2\!\!$\\

\hline
  0.9
& $\!\!$0.77$\pm$0.02(G)$\!\!$
& $\!\!$1.60$\pm$0.06(G)$\!\!$
& $\!\!$105/75$\!\!$
& $\!\!$1.69$\pm$0.04(G)
& $\!\!$1.44$\pm$0.18(E)
& $\!\!$120$\!\!$\\
  7.0
& $\!\!$2.28$\pm$0.01(E)$\!\!$
& $\!\!$0.68$\pm$0.03(E)$\!\!$
& $\!\!$692/75$\!\!$
& $\!\!$3.63$\pm$0.01(E)$\!\!$
& $\!\!$1.33$\pm$0.11(E)$\!\!$
& $\!\!$557$\!\!$\\
\hline
\end{tabular}
\vspace{1mm}
\begin{tabular}{c|ccccc}
\hline
$\!\!\sqrt s\!\!$ 
& \multicolumn{5}{c}{Eq.~(\ref{eq11})\quad CF$_{\rm II}$ ($\lambda_1$ and $\lambda_2$ : free)}\\
$\!\!\!\!$\scriptsize{[TeV]}$\!\!\!\!$  & $R_1$ [fm] & $\lambda_1$ & $R_2$ [fm] & $\lambda_2$ & $\chi^2$\\
\hline
0.9
& 0.87$\pm$0.03 (G)
& 0.26$\pm$0.02
& 2.8$\pm$0.3 (G)
& 0.47$\pm$0.07
& 79.8\\
7.0
& 1.85$\pm$0.02 (E)
& 0.59$\pm$0.01
& 3.5$\pm$0.1 (G)
& 0.28$\pm$0.01
& 466\\
\hline
& \multicolumn{5}{c}{Eq.(\ref{eq12}) with T-NBD \cite{Biyajima:2019} ($\lambda_1$ and $\lambda_2$ : fixed)}\\
\hline
0.9
& 0.85$\pm$0.02 (G)
& 0.39
& 2.55$\pm$0.10 (G)
& 0.24
& 81\\

7.0
& 1.87$\pm$0.01 (E)
& 0.59 
& 3.05$\pm$0.05 (G) 
& 0.18 
& 519\\
\hline
\end{tabular}
\end{table}

\section{\label{sec4}Concluding remarks}
\begin{figure}[h]
  \def\@captype{table}
  \begin{minipage}[c]{.48\textwidth}
  \centering
  \includegraphics[width=1.0\columnwidth]{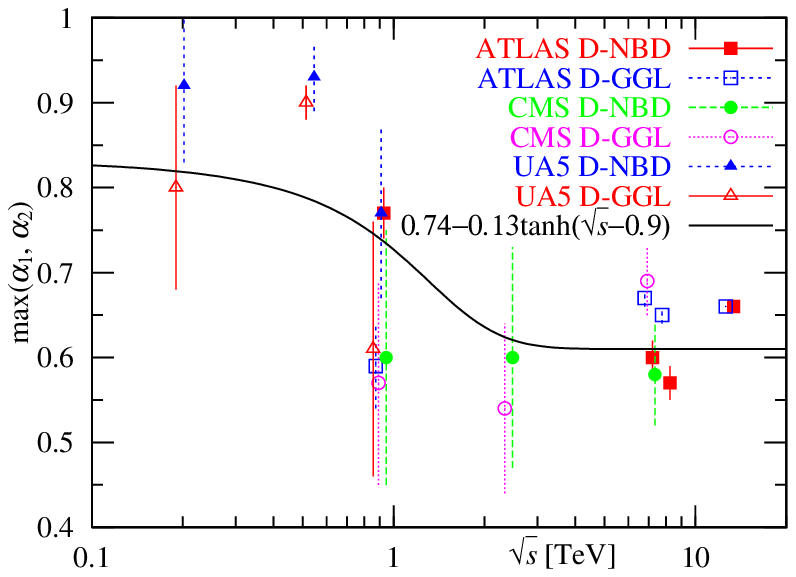}
  \caption{\label{fig3}The energy dependence of the $(\alpha_1,\,\alpha_2)$ parameters in Eqs.~(\ref{eq2}) and (\ref{eq5}).}
  \end{minipage}
  \hfill
  \begin{minipage}[t]{.48\textwidth}
\vspace{-35mm}
  \centering
  \includegraphics[width=1.0\columnwidth]{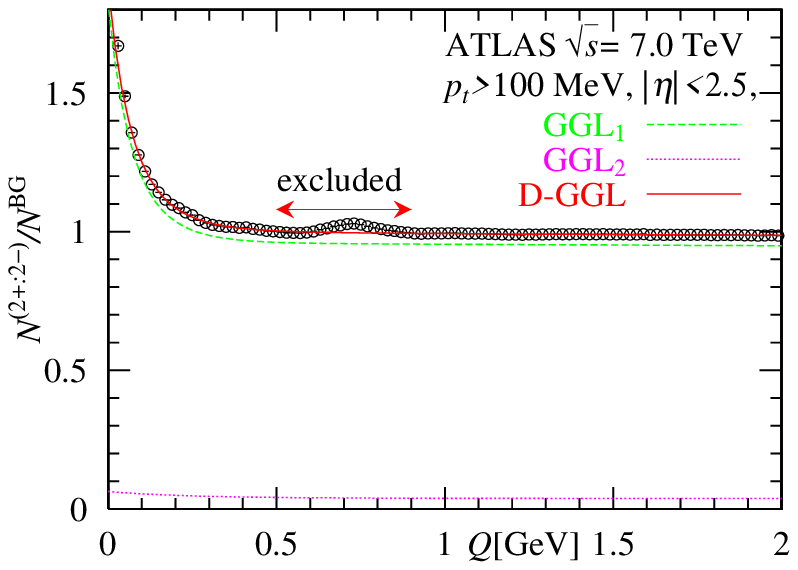}
  \caption{\label{fig4}Analyses of ATLAS data on the BEC at 7.0 TeV ($p_t>100$ MeV, $|\eta| < 2.5$, $n_{\rm ch} \ge 2$) using Eq.~(\ref{eq8}).}
  \end{minipage}
\vspace{-2mm}
\end{figure}

{\bf 1) } From the analysis of MD with $|\eta| < $2.5--3.0, we obtained the energy dependence of the weight factors $(\alpha_1,\,\alpha_2)$ (Fig.~\ref{fig3}). This behavior denotes the degree of the KNO scaling violation.

\noindent
{\bf 2) } The formula based on $N_{\rm II}^{\rm BG}$, i.e., Eq.~(\ref{eq8}), explains the BEC data more reasonably than the formula based on $N_{\rm I}^{\rm BG}$, i.e., Eq.~(\ref{eq7}) (Table \ref{tab2}). At present, the second denominator seems to be available for the study on the BEC. Detailed calculations will be provided elsewhere \cite{Biyajima:2019}.\\

\noindent
{\bf 3) } In 1986, the violation of KNO scaling was discovered by UA5 Collaboration. To explain this phenomenon, the UA5 collaboration proposed the D-NBD, i.e., Eq.~(\ref{eq2}). Moreover, in experiments at LHC energies, a remarkable violation of the KNO scaling has been found. In such situation, we proposed the unified analyses of MD and BEC at the LHC by making use of D-GGL as well as D-NBD. For BEC, we have derived Eqs.~(\ref{eq7}) and (\ref{eq8}). They are unified as follows:
\begin{eqnarray}
{\rm CF_{II}} = 1.0 + \lambda_1 E_{\rm BE_1}^2+\lambda_2 E_{\rm BE_2}^2,
\label{eq11}
\end{eqnarray}
where $\lambda_2$ is the second degree of coherence. Eq.~(\ref{eq11}) can be named a second conventional formula as CF$_{\rm II}$. Our result by Eq.~(\ref{eq11}) is also given in Table \ref{tab2}.\\

\noindent
{\bf 4) } Analyses of ATLAS BEC by the triple-NBD (T-NBD) \cite{Biyajima:2019} are added. The T-NBD is the extensive formula of Eq.~(\ref{eq2}):
\begin{eqnarray}
  P_{\rm T\mathchar`-NBD}(n,\, \langle n\rangle) = \sum_{i=1}^3 \alpha_i P_{{\rm NBD}_i}(n,\,k_i,\,\langle n_i\rangle).
\label{eq12}
\end{eqnarray}
For T-NBD, see also \cite{Zborovsky:2013tla}. The similarities with results by CF$_{\rm II}$ seem to be better those by D-NBD and D-GGL: Results by CF$_{\rm II}$ and those by Eq.~(\ref{eq12}) with the extensive formula of Eq.~(\ref{eq8}) are almost the  same, except for $\lambda$'s and $\chi^2$s.\\

\noindent
{\it Acknowledgments. } T.M. would like to appreciate the special budget given by Pres. Y. Hayashi. M.B. would like to thank the colleagues at the Department of Physics in Shinshu University for their kindness.



\begin{thebibliography}{99}
\bibitem{Aad:2010ac}
  G.~Aad {\it et al.} [ATLAS Collaboration],
  New J.\ Phys.\  {\bf 13} (2011) 053033.

\bibitem{Aaboud:2016itf}
  M.~Aaboud {\it et al.} [ATLAS Collaboration],
  Eur.\ Phys.\ J.\ C {\bf 76} (2016) no.9, 502.

\bibitem{Khachatryan:2010nk}
  V.~Khachatryan {\it et al.} [CMS Collaboration],
  JHEP {\bf 1101} (2011) 079.

\bibitem{Giovannini:1998zb} 
  A.~Giovannini and R.~Ugoccioni,
  Phys.\ Rev.\ D {\bf 59} (1999) 094020.

\bibitem{Ghosh:2012xh}
  P.~Ghosh,
  Phys.\ Rev.\ D {\bf 85} (2012) 054017.

\bibitem{Koba:1972ng} 
  Z.~Koba, H.~B.~Nielsen and P.~Olesen,
  Nucl.\ Phys.\ B {\bf 40} (1972) 317.

\bibitem{Fuglesang:1989st} 
  C.~Fuglesang,
  La Thuile Multiparticle Dynamics 1989 (1989) 193-210 (World Scientific, 1990).

\bibitem{Alner:1985zc}
  G.~J.~Alner {\it et al.} [UA5 Collaboration],
  Phys.\ Lett.\  {\bf 160B} (1985) 193.

\bibitem{Ansorge:1988kn} 
  R.~E.~Ansorge {\it et al.} [UA5 Collaboration],
  Z.\ Phys.\ C {\bf 43}, 357 (1989).

\bibitem{Aamodt:2010pp}
  K.~Aamodt {\it et al.} [ALICE Collaboration],
  Eur.\ Phys.\ J.\ C {\bf 68} (2010) 345.

\bibitem{Mizoguchi:2010vc}
  T.~Mizoguchi and M.~Biyajima,
  Eur.\ Phys.\ J.\ C {\bf 70} (2010) 1061.

\bibitem{GrosseOetringhaus:2009kz}
  J.~F.~Grosse-Oetringhaus and K.~Reygers,
  J.\ Phys.\ G {\bf 37} (2010) 083001.

\bibitem{Navin:2010kk}
  S.~Navin,
  ``Diffraction in Pythia,''
  LUTP-09-23
  [arXiv:1005.3894 [hep-ph]].

\bibitem{Biyajima:2003nv}
  M.~Biyajima {\it et al.},
  Prog.\ Theor.\ Phys.\ Suppl.\  {\bf 153} (2004) 344.

\bibitem{Biyajima:1978cz}
  M.~Biyajima, O.~Miyamura and T.~Nakai,
  in Proc. Multiparticle Dynamics (Research Inst. for Fundamental Physics, Kyoto Univ., 1978), p. 139

\bibitem{Biyajima:1979ak}
  M.~Biyajima,
  Phys.\ Lett.\ B {\bf 92} (1980) 193.

\bibitem{Biyajima:1990ku}
  M.~Biyajima, A.~Bartl, T.~Mizoguchi, O.~Terazawa and N.~Suzuki,
  Prog.\ Theor.\ Phys.\  {\bf 84} (1990) 931.

\bibitem{Kittel:2005fu}
  W.~Kittel and E.~A.~De Wolf,
  ``Soft multihadron dynamics,''
  (World Scientific, Singapore, 2005).

\bibitem{Biyajima:2018abe}
  M.~Biyajima and T.~Mizoguchi,
  Eur.\ Phys.\ J.\ A {\bf 54} (2018) no.6, 105.

\bibitem{Aad:2015sja}
  G.~Aad {\it et al.} [ATLAS Collaboration],
  Eur.\ Phys.\ J.\ C {\bf 75} (2015) no.10, 466.

\bibitem{Biyajima:2019}
  M.~Biyajima and T.~Mizoguchi,
  ``Description of the Bose--Einstein Correlation at the LHC using Double and Triple Probability Distributions,''
  in preparation.

\bibitem{Zborovsky:2013tla}
  I.~Zborovsky,
  J.\ Phys.\ G {\bf 40} (2013) 055005.

\end{thebibliography}
\end{document}